\documentclass[A4paper,10pt,conference]{ifacconf}

\usepackage{amsmath} 
\usepackage{amssymb}  
\usepackage{mathtools}
\usepackage{graphicx}
\usepackage{natbib}        
\usepackage{caption,subcaption}
\usepackage{color}
\usepackage{enumerate}
\usepackage{epstopdf}
\usepackage{dsfont}
\usepackage[font=small,skip=4pt]{caption}

\usepackage{pgfplots}
\pgfplotsset{width=10cm, compat=1.9, legend style={font=\footnotesize}}
\tikzset{%
	dot/.style={circle, fill=black, minimum size=4pt, inner sep=0pt, outer sep=-1pt},
	hdot/.style={circle, fill=white, minimum size=4pt, inner sep=0pt, outer sep=-1pt},
}
\newlength\fheight 
\newlength\fwidth 
\usepackage{circuitikz}
\usepgflibrary{arrows}

\usepackage{tikz}
\usetikzlibrary{calc, backgrounds}
\usetikzlibrary{shapes.geometric, arrows}

\tikzstyle{process}  = [rectangle, rounded corners, minimum width=2cm, minimum height=1cm,text centered, text width=2cm, draw=black, fill=red!20]

\tikzstyle{decision} = [rectangle, minimum width=2cm, minimum height=1cm, text centered, text width=2.5cm, draw=black, fill=orange!20]

\tikzstyle{scenario} = [rectangle, rounded corners, minimum width=2cm, minimum height=0.5cm, text centered, text width=3cm, draw=black, fill=pink!20]

\tikzstyle{arrow}= [thick, ->, >=stealth]

\tikzstyle{system} = [rectangle, rounded corners, minimum width=4cm, minimum height=4cm, text centered, text width=4cm, draw=black, fill=red!40,fill opacity = 0.2]

\tikzstyle{controller} = [rectangle,rounded corners,minimum width=1cm,minimum height = 1cm,text centered,text width=1cm,draw=black,fill=green!50,fill opacity = 0.2]

\tikzstyle{detector} = [rectangle,rounded corners,minimum width=1cm,minimum height = 1cm,text centered,text width=1cm,draw=black,fill=yellow!50,fill opacity = 0.2]

\newtheorem{proposition}{Proposition}

\newtheorem{assumption}{Assumption}
\newtheorem{remark}{Remark}
\newtheorem{problem}{Problem}

\newcommand{\Ncal}{{\mathcal{N}}}
\newcommand{\Gcal}{{\mathcal{G}}}
\newcommand{\Vcal}{{\mathcal{V}}}
\newcommand{\Ecal}{{\mathcal{E}}}
\newcommand{\Wcal}{{\mathcal{W}}}
\newcommand{\Acal}{{\mathcal{A}}}
\newcommand{\Lcal}{{\mathcal{L}}}
\newcommand{\Dcal}{{\mathcal{D}}}
\newcommand{\Scal}{{\mathcal{S}}}
\newcommand{\Ccal}{{\mathcal{C}}}
\newcommand{\RR}{{\mathbb{R}}}
\newcommand{\KK}{{\mathbb{K}}}
\newcommand{\I}{{\mathbf{I}}}
\newcommand{\1}{{\mathbf{1}}}
\newcommand{\0}{{\mathbf{0}}}

\newcommand{\commentout}[1]{}

\begin{document}
	
	\begin{frontmatter}
		
		\title{Consensusability of linear interconnected multi-agent systems
		}
		
		\thanks[footnoteinfo]{This work has been supported by the Swiss National Science Foundation	under the COFLEX project (grant number 200021\_169906).}

		\author[First]			{Mustafa S. Turan,} 
		\author[First]			{Liang Xu,} 
		\author[First]			{Giancarlo Ferrari-Trecate}
		
		\address[First]{Dependable Control and Decision Group (DECODE), \\
			Laboratoire d'Automatique, \\ 
			\'Ecole Polytechnique F\'ed\'erale de Lausanne (EPFL), Switzerland. (e-mails: \{mustafa.turan, liang.xu, giancarlo.ferraritrecate\}@epfl.ch)}

		\begin{abstract}
			Consensusability is an important property for many multi-agent systems (MASs) as it implies the existence of networked controllers driving the states of MAS subsystems to the same value. Consensusability is of interest even when the MAS subsystems are physically coupled, which is the case for real-world systems such as power networks. In this paper, we study necessary and sufficient conditions for the consensusability of linear interconnected MASs. These conditions are given in terms of the parameters of the subsystem matrices, as well as the eigenvalues of the physical and communication graph Laplacians. Our results show that weak coupling between subsystems
			and fast information diffusion rates in the physical and communication graphs favor consensusability. Technical results are verified through computer simulations. 
		\end{abstract}
		
	\end{frontmatter}

	\section{Introduction}\label{ch:Intro}
	\subsection{Motivation and state of the art}
	
	In the past decades, results on multi-agent systems (MASs) found applications in several fields,  ranging from biological networks to cooperative robotics and power systems.
	Consensus, i.e., the achievement of an agreement on a quantity of interest for all agents, is one of the most interesting collective behaviours for MASs. 
	
	Consensusability of MASs amounts to assessing whether there exists a distributed protocol such that the MAS can achieve consensus.
	The consensusability problem for linear MASs has been extensively studied in the past decades.
	For example, \cite{MaCuiqin2010TAC} show that for continuous-time linear MASs, if the dynamics of each agent is controllable and the communication topology is connected, the MAS can reach consensus.
	~\cite{you2011} show that for discrete-time linear MASs, the unstable eigenvalues of the agent state matrix should satisfy certain conditions depending on the eigenvalues of the communication Laplacian to guarantee consensus.
	For consensus in MASs with switching communication topologies,~\cite{WangXingping2018TAC} develop an approach based on the Lyapunov exponent of agent dynamics and a suitably defined synchronizability exponent for the switching topology.
	More specifically, they show that if the Lyapunov exponent is less than the synchronizability exponent, the MAS can achieve consensus.
	These results reveal connections between consensus, the agent dynamics and the switching topology.
	Furthermore,~\cite{XuLiang2018Automatica} study consensusability problems over random communication channels among agents.
	They provide a consensusability condition based on the statistics of the communication channel, the eigenvalues of the communication Laplacian and the instability degree of the agent dynamics.
	~\cite{LiLin2019Automatica}, \cite{Zhengjianying2019TAC} and \cite{XuJuanjuan2019TAC} consider time delay in communication channels among agents and derive upper bounds on communication delay to guarantee consensus.
	All the above research works provide analytic characterizations of linear MASs in order to achieve consensus with linear consensus protocols.
	
	However, the above research works assume that the agents have decoupled dynamics, which is not the case in some applications.
	For example, in power networks and microgrids, each node is physically interconnected with other nodes and consequently their dynamics are coupled~(\cite{guerrero2010}).
	Therefore, in this paper, we study the consensusability problem for linear \emph{interconnected} MASs (LIMASs) and try to reveal how it is affected by physical couplings among agents.
	Consensus for LIMASs has been studied by \cite{yangzhou2015consensus} and \cite{cheng2015leader}.
	\cite{yangzhou2015consensus} investigate consensusability of LIMASs with linear protocols.
	They show that the consensusability is equivalent to the stability of a matrix constructed from the agent dynamics and the consensus protocol.
	Moreover, an LMI-based design procedure is provided for the consensus protocol.
	~\cite{cheng2015leader} consider leader-follower tracking problems for LIMASs.
	Interactions between agents are treated as dynamic uncertainties and are described in terms of integral quadratic constraints.
	A leader-follower tracking control protocol is proposed and sufficient conditions to guarantee that the followers track the leader's state are obtained in terms of the feasibility of LMIs.
	However, \cite{yangzhou2015consensus} and \cite{cheng2015leader} only give consensus conditions in terms of the feasibility of certain LMIs, which provide little insight into how physical couplings affect consensus.

	\subsection{Objectives and contributions}\label{subch:Objectives}
	
	In this paper, we are interested in studying how the coupling among agents affect the consensus problem.
	For this purpose, we provide analytical characterizations of the consensusability of LIMASs.
	Specifically, for general systems, we show that linear consensus protocol design is equivalent to a control design problem with structural constraints, which is intractable.
	With the help of technical assumptions introduced to simplify the problem, we show that the consensusability of the LIMAS is equivalent to a simultaneous stabilizability problem.
	Then we derive first a sufficient condition and then a necessary condition for consensusability.
	For LIMAS with scalar agent dynamics, we remove the requirement of the technical assumptions by directly analyzing the eigenvalues of the state matrix of the consensus error dynamics.
	A sufficient consensus condition is further provided.
	All the derived results clearly illustrate how the existence of a linear distributed consensus protocol for LIMASs is influenced by the interplay between the coupling matrix, the coupling topology, the agent dynamics and the control topology.
	
	This paper is organized as follows.
	The problem formulation is given in Section~\ref{sec:ProblemFormulation}.
	The consensus conditions are derived in Section~\ref{sec:Consensusability}.
	Simulations are provided in Section~\ref{sec:SimulationResults} and some concluding remarks are provided in Section~\ref{sec:ConclusionAndFuturePerspectives}.

	\subsection{Notation}
	In the paper, the operator $|\cdot|$ applied to a set determines its cardinality, while used with matrices or vectors it defines their component-by-component absolute value.
	The operator $\|\cdot\|$ is used to define the matrix norm.
	We use $A>0$ ($A\geq 0$) for denoting that the specific matrix $A$ is positive (semi-) definite.
	The symbol $\otimes$ represents the Kronecker product.
	$\1_n \in \RR^n$ and $\0_n \in \RR^n$ represent column vectors with all elements equal to one and zero, respectively.
	
	\subsection{Preliminaries on algebraic graph theory}
	
	An undirected weighted graph of $N$ nodes is defined as $\Gcal = \left(\Vcal, \Wcal, \Ecal\right)$, where $\Vcal = \{1,2,\dots,N\}$ is the set of nodes, $\Ecal \subseteq \Vcal \times \Vcal$ is the set of edges with $\left|\Ecal\right|=M$, and $\Wcal \in \RR^{M\times M}$ is a diagonal matrix with the weight of the corresponding edge in each diagonal entry. For each node $i$, the set of its neighbors is defined as $\Ncal_i = \{j|(i,j)\in\Ecal\}$. A path $p_{ij}$ is defined as a sequence of consecutive edges such that every edge in the sequence is in $\Ecal$, the first edge starts from node $i$, and the last edge ends in node $j$. The adjacency matrix of $\Gcal$ is defined as $\Acal = \left[a_{ij}\right]_{N\times N}$, where $a_{ij} = 0$ if $\left(i,j\right)\notin \Ecal$ or $i=j$ and $a_{ij} >0$ is the positive weight of edge $\left(i,j\right) \in \Ecal$. The degree of a node $i$ is defined as $d_i = \sum_{j\in\Ncal_i}a_{ij}$ along with the degree matrix $\Dcal = diag\left(d_1,d_2,\dots,d_N\right)$. The Laplacian matrix of $\Gcal$ is given by $\Lcal = \Dcal - \Acal$. An undirected graph is connected if there exists a path from every node $i$ to every other node $j$ for all $i,j\in\Vcal$. 
	
	\section{Problem Formulation}\label{sec:ProblemFormulation}
	
	\begin{figure}
		\centering
		{\tiny
			\ctikzset{bipoles/length=0.8cm}
			\tikzstyle{every node}=[minimum size=.8cm, inner sep=.8, line width=0.5pt]	
			\begin{circuitikz}[american currents, scale=0.73, line width=1pt]
				\draw (1,1) node(s1) [circle, draw=black, double,
				fill=black!20] {$\Scal_1$};
				\draw (7,-.1) node(s4)  [circle, draw=black, double,fill=black!20] {$\Scal_4$};
				\draw (3,-.1) node(s3)  [circle, draw=black, double, fill=black!20] {$\Scal_3$};
				\draw (5,1) node(s2)  [circle, draw=black, double,  fill=black!20] {$\Scal_2$};

				\draw[latex-latex, blue] (s1) to (s2);
				\draw[latex-latex, blue] (s2)  to (s4);  
				\draw[latex-latex, blue] (s4) to (s3);
				\draw[latex-latex, blue] (s1) to (s3);

				\draw (1,1-3.5) node(c1) [circle, draw=black, double,
				fill=black!20] {$\Ccal_1$};
				\draw (7,-.1-3.5) node(c4)  [circle, draw=black, double,fill=black!20] {$\Ccal_4$};
				\draw (3,-.1-3.5) node(c3)  [circle, draw=black, double, fill=black!20] {$\Ccal_3$};
				\draw (5,1-3.5) node(c2)  [circle, draw=black, double,  fill=black!20] {$\Ccal_2$};
				
				\draw[dashed, gray] (s1) to(c1);
				\draw[dashed, gray] (s2) to (c2);
				\draw[dashed, gray] (s3) to (c3);
				\draw[dashed, gray] (s4) to (c4);
				
				\draw[latex-, black] (8,1) to [bend right=10] (6.3,0.6) ;		
				\draw[latex-, black] (8,1-3.5) to [bend right=10] (6.3,0.6-3.5) ;		
				
				\node[anchor=west,text width=1.5cm] (note1) at (8,1)
				{\textbf{Physical Layer}};
				
				\node[anchor=west,text width=1.5cm] (note2) at (8,1-3.5)
				{\textbf{Control Layer}};
				
				\draw[latex-latex, red] (c3) to (c2);
				\draw[latex-latex, red] (c2) to (c1);
				\draw[latex-latex, red] (c4) to (c2);
		\end{circuitikz}}
		\caption{Graph of a LIMAS. The blue arrows represent the physical couplings between the subsystems $\Scal_i$, dashed gray lines indicate corresponding elements in the physical layer and control layer, and the red arrows represent the communication channels between the controllers $\Ccal_i$. 
		}
		\label{fig:LIMAS_struct}
	\end{figure}
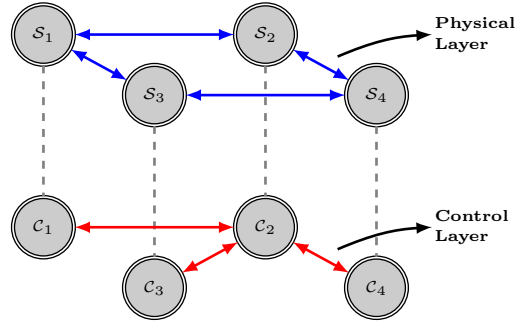
	
	We consider the problem of consensusability for linear interconnected multi-agent systems (LIMASs). A LIMAS is defined by two undirected weighted graphs: a physical graph $\Gcal_p = \left(\Vcal,\Wcal_p,\Ecal_p\right)$ representing the physical coupling between subsystems that are influenced by each other's dynamics and a communication graph $\Gcal_c = \left(\Vcal,\Wcal_c,\Ecal_c\right)$ representing the cyber coupling between subsystems, introduced by the communication network for information exchange between subsystems. 
	
	\begin{remark}
		Typical examples of a LIMAS are DC microgrids, where distributed generation units are physically coupled through electric lines and communication networks are used for obtaining global coordinated bahaviors of electric variables, such as current sharing and voltage balancing (\cite{tucci2018}).
	\end{remark}
	
	In the scope of this work, we assume that the communication graph $\Gcal_c$ is connected, as this is necessary to reach consensus. For each subsystem $\Scal_i$, the set of its neighbors in the physical graph $\Gcal_p$ is denoted by $\Ncal_i^p\triangleq\{j|(i,j)\in\Ecal_p\}$, whereas, similarly, the set of its neighbors in the communication graph $\Gcal_c$ is denoted by $\Ncal_i^c\triangleq\{j|(i,j)\in\Ecal_c\}$. Figure~\ref{fig:LIMAS_struct} presents an example LIMAS with $N=4$ subsystems, where physical graph edges are shown with blue arrows and communication graph edges are shown with red arrows. In particular, a LIMAS consists of $N$ subsystems where each subsystem $\Scal_i$ has the following dynamics:
	\begin{equation}\label{eq:Si_dyn}
	x_i^+ = Ax_i + A_p\sum_{j\in\Ncal_i^p}a_{ij}(x_j-x_i) + Bu_i,
	\end{equation}
	where $x_i \in \RR^n$ are the states of the subsystem, $u_i \in \RR$ are the scalar control inputs to the subsystem, $A \in \RR^{n\times n}$ is the state matrix identical to all subsystems, $a_{ij}=a_{ji} \in \RR_{>0}$ are the symmetrical physical interconnection weights between subsystems $i$ and its neighbors in the physical graph $\Gcal_p$, $A_p \in \RR^{n\times n}$ is a matrix determining the type of physical coupling, and $B \in \RR^{n\times 1}$ is the control matrix identical to all subsystems. 
	
	In this paper, we are interested in \textit{networked} controllers producing scalar control inputs $u_i$ by utilizing the information from neighbors. More specifically, consensus-based controllers are considered:
	\begin{equation}\label{eq:Si_contr}
	u_i = K\sum_{j\in\Ncal_i^c}b_{ij}\left(x_i-x_j\right),
	\end{equation}
	where $K\in\RR^{1\times n}$ is a control gain that is common to all subsystems and $b_{ij}=b_{ji}\in \RR_{>0}$ denote the symmetrical weights of the communication graph $\Gcal_c$. This particular class of controllers is used to make the states of every subsystem converge to each other, i.e., achieve that $|x_i-x_j|\rightarrow0$ for all $i,j\in \Vcal$. By combining \eqref{eq:Si_dyn} and \eqref{eq:Si_contr}, the overall dynamics of the whole LIMAS can be written as:
	\begin{equation}\label{eq:LIMAS_dyn}
	x^+ = \left(\I_N \otimes A - \Lcal_p\otimes A_p + \Lcal_c\otimes BK\right)x,
	\end{equation}
	where $x = \left[x_1^{\top},\dots,x_N^{\top}\right]^{\top} \in \RR^{Nn}$ is the cumulative state and $\Lcal_p$, $\Lcal_c$ are the Laplacian matrices associated with $\Gcal_p$ and $\Gcal_c$, respectively. Note that, in the field of consensus for MASs, it is common to assume identical subsystem dynamics and feedback gains as in \eqref{eq:Si_dyn},\eqref{eq:Si_contr}, since application areas of interest, such as vehicle formation, inherently consider a \textit{swarm} of identical agents achieving a collective task~(\cite{fax2004information, ren2007information}). Moreover, using a single static feedback gain for all agents would also alleviate the burden of designing different consensus gains for each agent, which is critical in deploying large scale multi-agent systems. We are interested in the following problem:
\begin{problem}[Consensusability]\label{prob:Consensusability}
	Given the dynamics of LIMAS in \eqref{eq:LIMAS_dyn}, provide conditions for the existence of a static feedback gain $K \in \RR^{1\times n}$ such that the states of the subsystems converge to a global consensus state, i.e., the states $x_i(t)$ satisfy
	\begin{equation}\label{eq:consensusdefinition}
		\lim\limits_{t\rightarrow\infty}|x_i(t)-\bar{x}| = 0, \quad \forall i \in \Vcal,
	\end{equation}
	for some $\bar{x} \in \RR^{n}$.
\end{problem}

	\begin{remark}
		We note that the formulation in~\eqref{eq:LIMAS_dyn} matches models commonly used in the literature on consensusability of linear multi-agent systems~(\cite{you2011}), with the difference of the physical coupling term $\Lcal_p\otimes A_p$. Because of this term, the method in \cite{you2011} cannot be used to analyze the consensusability of \eqref{eq:LIMAS_dyn}.
	\end{remark}
	
	Since the subsystem states need to converge to the same value, one can define the average of the cumulative state $x$ as $$\bar{x} \triangleq \frac{1}{N} \sum_{i=1}^Nx_i = \frac{1}{N}\left(\1_N^{\top}\otimes\I_n\right)x,$$ and analyze the evolution of the deviation from this consensus state:
	\begin{equation}\label{eq:delta_def}
	\begin{split}
	\delta &\triangleq \left[x_1^{\top}-\bar{x}^{\top},\dots,x_N^{\top}-\bar{x}^{\top}\right]^{\top} = x-\1_N\otimes\bar{x} \\
	&= \left(\I_{Nn}-\frac{1}{N}\left(\1_N\1_N^{\top}\right)\otimes\I_n\right)x \\
	&= \left(\left(\I_N-\frac{1}{N}\1_N\1_N^{\top}\right)\otimes\I_n\right)x
	\end{split}.
	\end{equation}
	
	In particular, \eqref{eq:consensusdefinition} is equivalent to $\lim \limits_{t\rightarrow\infty}|\delta(t)| = 0$ for all initial conditions $\delta(0)$. For this purpose, we study the dynamics of $\delta$, which is derived from \eqref{eq:LIMAS_dyn} and \eqref{eq:delta_def} as:
	
	\begin{equation}\label{eq:delta_dyn}
	\begin{split}
	\delta^+ &=  \left(\left(\I_N-\frac{1}{N}\1_N\1_N^{\top}\right)\otimes\I_n\right)x^+ \\
	&= \left(\I_N \otimes A - \Lcal_p\otimes A_p + \Lcal_c\otimes BK\right)\delta
	\end{split}.
	\end{equation}
	
	As the consensusability of~\eqref{eq:LIMAS_dyn} is equivalent to the stabilizability of~\eqref{eq:delta_dyn}, Problem~\ref{prob:Consensusability} is equivalent to the following problem:
	
	\begin{problem}[Stabilizability] \label{prob:stabilizability}
		Provide conditions for the existence of a static feedback gain $K\in\RR^{1\times n}$ such that $\left(\I_N \otimes A - \Lcal_p\otimes A_p + \Lcal_c\otimes BK\right)$ is Schur stable, i.e., its eigenvalues are within the unit circle in the complex plane.
	\end{problem}
	
	\section{Consensusability}\label{sec:Consensusability}
	
	In the sequel, we analyze separately consensusability for subsystems of order $n>1$ and $n=1$. This is because the former case is more difficult to analyze without further assumptions. Instead, the latter case with scalar subsystem dynamics can be studied without restrictive assumptions and gives important insight into the solution of Problem~\ref{prob:stabilizability}.

	\subsection{Subsystems of order $n>1$}\label{subsec:VectorDynamics}
	
	Problem~\ref{prob:stabilizability} can be seen as the problem of designing controllers with structure constraints. Indeed, the aim is to make the matrix $$\I_N\otimes A-\Lcal_p\otimes A_p+\left(\Lcal_c\otimes B\right)\underbrace{\left(\I_N\otimes K\right)}_{\tilde{K}}$$ Schur stable by designing the state feedback $\tilde{K}$ with a sparsity constraint, i.e., a decentralized controller. Prior work in literature show that this problem is difficult to tackle without focusing on special system structures~(\cite{blondel2000,rotkowitz2005}). Therefore, to facilitate the analysis, we introduce the following technical assumption. 
	
	\begin{assumption}\label{ass:lapl_commute}
		The Laplacian matrices $\Lcal_p$ and $\Lcal_c$ commute. 
	\end{assumption}
	
\begin{remark}\label{rem:commutingLaplacians}
	Note that Assumption~\ref{ass:lapl_commute} is verified when the two Laplacians are equal to each other up to a scaling constant, i.e., $\Lcal_p=\beta \Lcal_c$, $\beta\in\RR$. Moreover, two Laplacians commute also when one of them is the Laplacian of a complete graph with uniform edge weights. Nevertheless, necessary and sufficient conditions for two generic Laplacians to commute are not yet available in the literature.
\end{remark}
	
	It is well known that, under Assumption~\ref{ass:lapl_commute}, one can simultaneously diagonalize the two Laplacians $\Lcal_p$ and $\Lcal_c$~(\cite{horn1985}). Therefore, it is possible to find a unitary matrix $\Phi = \left[\1_N/\sqrt{N},\phi_2,\dots,\phi_N\right]$ such that $\Phi^{\top}\Lcal_p\Phi = \Lambda_p = diag\left(0,\lambda_{p,2},\dots,\lambda_{p,N}\right)$ and $\Phi^{\top}\Lcal_c\Phi = \Lambda_c = diag\left(0,\lambda_{c,2},\dots,\lambda_{c,N}\right)$ are diagonal matrices that collect the eigenvalues of $\Lcal_p$ and $\Lcal_c$ in their diagonals, respectively. Note that the entries of $\Lambda_p$ and $\Lambda_c$ are not necessarily ordered by their magnitude, with the exception of the first diagonal entries of both matrices, which are $0$. Next, we denote the largest and second smallest eigenvalues of $\Lcal_p$ as $\lambda_{p,max}\triangleq\max_{i\in\{2,\dots,N\}}\lambda_{p,i}$ and $\lambda_{p,min}\triangleq\min_{i\in\{2,\dots,N\}}\lambda_{p,i}$, respectively. For $\Lcal_c$, $\lambda_{c,max}$ and $\lambda_{c,min}$ are similarly defined.
	
	Using these definitions, the dynamics of $\tilde{\delta} = \left[\tilde{\delta}_1^{\top},\dots,\tilde{\delta}_N^{\top}\right]^{\top}$ $\triangleq \left(\Phi^{\top}\otimes\I_n\right) \delta$ can be written as:
	\begin{equation}\label{eq:deltatilde_dyn}
	\tilde{\delta}^+ = \left(\I_N\otimes A-\Lambda_p\otimes A_p+\Lambda_c\otimes BK\right)\tilde{\delta}.
	\end{equation}
	
	As the system matrices in \eqref{eq:delta_dyn} and \eqref{eq:deltatilde_dyn} are congruent to each other, their stabilizability properties are equivalent. We further note that, using \eqref{eq:delta_def}, one has $\tilde{\delta}_1 = \left(\frac{1}{\sqrt{N}}\1_N^{\top}\otimes\I_n\right)\delta = \0_n$.
	
	Note that, from Assumption~\ref{ass:lapl_commute}, the consensusability of \eqref{eq:LIMAS_dyn} is equivalent to the stabilizability of \eqref{eq:deltatilde_dyn}, i.e., existence of a $K\in\RR^{n\times m}$ such that $$\rho\left(A-\lambda_{p,i}A_p+\lambda_{c,i}BK\right)<1 \quad \forall i\in\{2,\dots,N\}.$$
	
	In order to show the existence of such a feedback gain $K$, it is required that each system $\tilde{\delta}_i$ for $i\in \{2,\dots,N\}$ is controllable. Therefore, we assume the following: 
	
	\begin{assumption}\label{ass:controllability}
		The pairs $(A-\lambda_{p,i}A_p,B)$ are controllable for all $i\in \{2,\dots,N\}$.
	\end{assumption}
	
	\begin{remark}
		We note that this assumption is required, as the controllability of all the pairs $(A-\lambda_{p,i}A_p,B)$ is not implied by that of the pair $(A,B)$ for general LIMASs. For instance, when $A_p=A$ and $\lambda_{p,i}=1$ is an eigenvalue of $\Lcal_p$, one has $A-\lambda_{p,i}A_p=0$ and the pair $(A-\lambda_{p,i}A_p,B)$ is uncontrollable unless $B$ is full rank. We also stress that the controllability of $(A-\lambda_{p,i}A_p,B)$ implies that of $(A-\lambda_{p,i}A_p,\lambda_{c,i}B)$, as $\lambda_{c,i}>0,$ $\forall i\in\{2,\dots,N\}$.
	\end{remark}
	
	Without further assumptions on the structures of $A$ and $A_p$, stabilizability of \eqref{eq:deltatilde_dyn} is the same as the problem of simultaneous stabilizability of a set of linear systems with a common static state feedback, given as $$x^+=(A_i+B_iK)x.$$ Despite a number of results on the simultaneous stabilzability of linear systems~(\cite{kabamba1991,vidyasagar1982,saadatjoo2009,blondel1994}), it is difficult to give conditions on the simultaneous stabilizability of more than two systems~(\cite{blondel1993}). Therefore, we analyze the consensusability problem in the following scenario.
	
	\begin{assumption}\label{ass:A_and_Ap}
		The physical interconnection matrix $A_p$ satisfies $A_p = \alpha A$, where $\alpha\in \RR$.
	\end{assumption}
	
	The following proposition presents a sufficient condition for consensusability of LIMASs under Assumption~\ref{ass:A_and_Ap}.
	
	\begin{proposition}\label{prop:consensusability_sufficient}
		Suppose that Assumptions~\ref{ass:lapl_commute}, \ref{ass:controllability}, and \ref{ass:A_and_Ap} hold. Then, the LIMAS in \eqref{eq:LIMAS_dyn} is consensusable if 
		\begin{equation}\label{eq:consensusability_sufficient_critical}
		\left(\frac{\max_{i,j}\frac{\alpha_i}{\lambda_{c,j}}-\min_{i,j}\frac{\alpha_i}{\lambda_{c,j}}}{2}\right)^2<\frac{\alpha_{min}^2-\alpha_{max}^2\sigma_c}{\lambda_{c,max}^2}
		\end{equation}
		where $\alpha_i\triangleq 1-\alpha\lambda_{p,i}$, $\alpha_{max} \triangleq \max_i|\alpha_i|$, $\alpha_{min} \triangleq \min_i|\alpha_i|$, and 
		\begin{equation}\label{eq:sigma_c}
		\sigma_c\triangleq \begin{cases}
		1-\frac{1}{\prod_{k}\left|\lambda_k^u\left(\alpha_{max}A\right)\right|^2},& \text{if } \rho\left(\alpha_{max}A\right) \geq 1\\
		0,              & \text{otherwise}
		\end{cases}
		\end{equation}
		with $\lambda^u(.)$ denoting the unstable eigenvalues of its argument. If the condition \eqref{eq:consensusability_sufficient_critical} is satisfied, the control gain $K$ is given by $K=-k(B^{\top}PB)^{-1}B^{\top}PA$ with $k=\frac{\min_{i,j}\frac{\alpha_i}{\lambda_{c,j}}+\max_{i,j}\frac{\alpha_i}{\lambda_{c,j}}}{2}$ and $P$ equal to the solution to the modified algebraic Riccati equation (MARE) given by~\eqref{eq:MARE}.
	\end{proposition}
	
	\begin{pf}
		Under Assumptions~\ref{ass:lapl_commute} and \ref{ass:A_and_Ap}, consensusability of \eqref{eq:LIMAS_dyn} is equivalent to the simultaneous stabilizability of 
		\begin{equation}\label{eq:deltatilde_dyn_assumptionAp}
			\tilde{\delta}_i^+ = \left(\underbrace{(1-\alpha\lambda_{p,i})}_{\alpha_i}A+\lambda_{c,i}BK\right)\tilde{\delta} \quad \forall i\in\{2,\dots,N\},
		\end{equation}
		which is equivalent to the existence of a $P_i> 0$ and $K$ such that $$\left(\alpha_iA+\lambda_{c,i}BK\right)^{\top}P_i\left(\alpha_iA+\lambda_{c,i}BK\right)-P_i<0,$$ for all $i\in\{2,\dots,N\}$. Letting $P_i=P$ for all $i$ and selecting a static feedback gain as $K = -k(B^{\top}PB)^{-1}B^{\top}PA$, one gets the following condition:
		\begin{equation}\label{eq:lyapunovcond}
			\begin{split}
				\alpha_i^2A^{\top}PA+&\left(\lambda_{c,i}^2k^2-2\alpha_i\lambda_{c,i}k\right)A^{\top}PB\left(B^{\top}PB\right)^{-1}B^{\top}PA \\
				&-P< 0
			\end{split},
		\end{equation}
		for all $i \in \{2,\dots,N\}$. Defining $\alpha_{max} = \max_i|\alpha_i|$, it can be seen that \eqref{eq:lyapunovcond} is satisfied for all $i\in\{2,\dots,N\}$ if 
		\begin{equation}\label{eq:lyapunovcond_max}
			\begin{split}
				\alpha_{max}^2A^{\top}PA+&\left(\lambda_{c,i}^2k^2-2\alpha_i\lambda_{c,i}k\right)A^{\top}PB\left(B^{\top}PB\right)^{-1}B^{\top}PA \\
				&-P< 0
			\end{split}
		\end{equation}
		is satisfied for all $i$. We note that, \eqref{eq:lyapunovcond_max}, for each $i$, can be written for the pair $(\bar{A},B)$ with $\bar{A}\triangleq \alpha_{max}A$:
		\begin{equation}\label{eq:MARE}
			\begin{split}
				\bar{A}^{\top}P\bar{A}-\sigma_i\bar{A}^{\top}PB\left(B^{\top}PB\right)^{-1}B^{\top}P\bar{A}-P< 0
			\end{split},
		\end{equation}
		where $\sigma_i\triangleq \frac{2\alpha_i\lambda_{c,i}k-\lambda_{c,i}^2k^2}{\alpha_{max}^2}$. We emphasize that, if $\bar{A}$ is stable, the systems in \eqref{eq:deltatilde_dyn_assumptionAp} are already stable with $K=0$, hence, simultaneously stabilizable, which corresponds to the solution of \eqref{eq:MARE} with $\sigma_i=0$ for all $i$. Proceeding with the case of unstable $\bar{A}$, we leverage the results in \cite{sinopoli2004} and \cite{schenato2007} to show that a critical value $\sigma_c \in [0,1)$ can be found as in \eqref{eq:sigma_c} such that there exists a $P> 0$ that solves \eqref{eq:MARE} for all $i \in \{2,\dots,N\}$ simultaneously if and only if $\min_i\sigma_i>\sigma_c$. Taking into consideration that $\sigma_i$ is a function of $k$, $\lambda_{p,i}$, and $\lambda_{c,i}$, along with the fact that $\lambda_{p,i}$s and $\lambda_{c,i}$s have no particular order, the condition on the solvability of \eqref{eq:MARE} transforms into the existence of a $k$ such that 
		\begin{equation}\label{eq:min_ij}
			\min_{i,j}\frac{2\alpha_i\lambda_{c,j}k-\lambda_{c,j}^2k^2}{\alpha_{max}^2}>\sigma_c
		\end{equation}
		holds. 
		Noticing that $2\alpha_i\lambda_{c,j}k-\lambda_{c,j}^2k^2= -\lambda_{c,j}^2(k-\frac{\alpha_i}{\lambda_{c,j}})^2+\alpha_i^2$, we get:
		\begin{equation*}
			\begin{split}
				\min_{i,j}2\alpha_i\lambda_{c,j}k-\lambda_{c,j}^2k^2 \geq -\max_{i,j}\lambda_{c,j}^2\left(k-\frac{\alpha_i}{\lambda_{c,j}}\right)^2+\alpha_{min}^2.
			\end{split}
		\end{equation*}
		Also exploiting the fact that $$\max_{i,j}\lambda_{c,j}^2\left(k-\frac{\alpha_i}{\lambda_{c,j}}\right)^2\leq \max_{j}\lambda_{c,j}^2\max_{i,j}\left(k-\frac{\alpha_i}{\lambda_{c,j}}\right)^2,$$ we can derive the following condition guaranteeing that \eqref{eq:min_ij} is satisfied:
		\begin{equation}\label{eq:minmax}
			\min_k \max_{i,j}\left(k-\frac{\alpha_i}{\lambda_{c,j}}\right)^2 <\frac{\alpha_{min}^2-\alpha_{max}^2\sigma_c}{\lambda_{c,max}^2}.
		\end{equation}
		Using that $\max_{i,j}(k-\frac{\alpha_i}{\lambda_{c,j}})^2 = (\max_{i,j}|k-\frac{\alpha_i}{\lambda_{c,j}}|)^2$ and defining the function $f(k) \triangleq \max_{i,j}|k-\frac{\alpha_i}{\lambda_{c,j}}| \geq 0$, it can be shown that $\arg \min_k f(k)^2 = \arg \min_kf(k)$. Moreover, thanks to the definition of $f(k)$, $$\arg \min_kf(k)=\frac{\min_{i,j}\frac{\alpha_i}{\lambda_{c,j}}+\max_{i,j}\frac{\alpha_i}{\lambda_{c,j}}}{2},$$ which leads to $$\min_k \max_{i,j}\left(k-\frac{\alpha_i}{\lambda_{c,j}}\right)^2=\left(\frac{\max_{i,j}\frac{\alpha_i}{\lambda_{c,j}}-\min_{i,j}\frac{\alpha_i}{\lambda_{c,j}}}{2}\right)^2,$$ concluding the proof.
		$\hfill \triangleleft$
	\end{pf}
	
	\begin{remark}\label{rmrk:consensusability_sufficient}
		We point out that condition~\eqref{eq:consensusability_sufficient_critical} can be satisfied only if its right hand side is positive. Therefore, it is necessary that $\alpha_{min}^2>\alpha_{max}^2\sigma_c$ for \eqref{eq:consensusability_sufficient_critical} to hold. Taking into account that $\sigma_c$ increases with $\alpha_{max}$, this is possible when $\alpha_{max}$ is \textit{small} and close to $\alpha_{min}$. Note that $\alpha_{max}$ takes small values for low values of $\alpha$, i.e., when the effect of physical coupling is not strong. Moreover, $\alpha_{min}$ and $\alpha_{max}$ are close to each other when $\alpha$ is small or $\lambda_{p,min}$ and $\lambda_{p,max}$ are close to each other. Also note that the left-hand side of the inequality in \eqref{eq:consensusability_sufficient_critical} decreases as the ratio $\frac{\lambda_{c,min}}{\lambda_{c,max}}$ increases, i.e., the eigenvalues of the communication graph get closer to each other. Since a bigger ratio $\frac{\lambda_{c,min}}{\lambda_{c,max}}$ signifies better synchronizability of a graph (\cite{you2011}), we conclude that condition \eqref{eq:consensusability_sufficient_critical} is easier to satisfy when the physical interconnection is weaker and both physical and communication graphs diffuse information more quickly, i.e., when they are closer to the complete graph. 
	\end{remark}
	
	We also provide the following result, which presents a necessary condition for the consensusability of the LIMAS in \eqref{eq:LIMAS_dyn}.
	
	\begin{proposition}\label{prop:consensusability_necessary}
		Suppose that Assumptions~\ref{ass:lapl_commute} and \ref{ass:controllability} hold and define $\gamma_c \triangleq \frac{\lambda_{c,max}}{\lambda_{c,min}}\geq1$. Then, a necessary condition for the stabilizability of the LIMAS in \eqref{eq:LIMAS_dyn} is:
		\begin{equation}\label{eq:consensusability_necessary}
		\left|\gamma_c\min_i|\det(A-\lambda_{p,i}A_p)|-\max_i|\det(A-\lambda_{p,i}A_p)|\right|<\gamma_c+1.  
		\end{equation}
	\end{proposition}
	
	\begin{pf}
		The proof is a modification of the proof of Lemma 3.1 in \cite{you2011}. We first show that, under Assumption~\ref{ass:controllability}, without loss of generality, each pair $(A-\lambda_{p,i}A_p,B)$ can be written in controllable canonical form
		\begin{equation}
			A-\lambda_{p,i}A_p = \begin{bmatrix}
				0 & 1 & 0 & \dots  \\
				\vdots & \ddots & \ddots & \ddots \\
				0 & \dots & 0 & 1 \\
				-a_{i,1} & -a_{i,2} & \dots & -a_{i,n}
			\end{bmatrix}, \quad B = \begin{bmatrix}
				0 \\
				\vdots \\
				0 \\
				1
			\end{bmatrix},
		\end{equation}
		where $|a_{i,1}| = |\det(A-\lambda_{p,i}A_p)|$. Given a simultaneously stabilizing feedback gain $K = \left[k_1, \dots, k_N\right]$, one can see that $|\det(A-\lambda_{p,i}A_p+\lambda_{c,i}BK)| = |a_{i,1}-\lambda_{c,i}k_1|$. As $K$ is selected to stabilize $(A-\lambda_{p,i}A_p,\lambda_{c,i}B)$ for all $i \in \{2,\dots,N\}$, $\rho(A-\lambda_{p,i}A_p+\lambda_{c,i}BK)<1$ and hence, 
		\begin{equation*}
			\begin{split}
				|\det(A-&\lambda_{p,i}A_p+\lambda_{c,i}BK)|\\
				&=\prod_{j}|\lambda_j\left(A-\lambda_{p,i}A_p+\lambda_{c,i}BK\right)|<1
			\end{split}.
		\end{equation*}
		This yields $|a_{i,1}-\lambda_{c,i}k_1|<1$, which can be further manipulated to give
		\begin{equation}
			\frac{|a_{i,1}|-1}{\lambda_{c,i}}<|k_1|< \frac{|a_{i,1}|+1}{\lambda_{c,i}}.
		\end{equation}
		This implies that $\bigcap_{i=2}^N\left(\frac{|a_{i,1}|-1}{\lambda_{c,i}},\frac{|a_{i,1}|+1}{\lambda_{c,i}}\right)\neq \emptyset$. Therefore, it holds that
		\begin{equation}\label{eq:a_i_bounds}
			\max_i\frac{|a_{i,1}|-1}{\lambda_{c,i}} < \min_i\frac{|a_{i,1}|+1}{\lambda_{c,i}}.
		\end{equation}
		At this point, we exploit the facts $\max_i\frac{|a_{i,1}|-1}{\lambda_{c,i}}\geq\max_i\frac{\min_j|a_{j,1}|-1}{\lambda_{c,i}}$ and $\min_i\frac{|a_{i,1}|+1}{\lambda_{c,i}}\leq\min_i\frac{\max_j|a_{j,1}|+1}{\lambda_{c,i}}$, as well as the definition of $\gamma_c$ to show that \eqref{eq:a_i_bounds} implies
		\begin{equation*}
			\gamma_c\min_i|\det(A-\lambda_{p,i}A_p)|-\max_i|\det(A-\lambda_{p,i}A_p)|<\gamma_c+1.  
		\end{equation*}
		Similarly, it also holds that $\max_i\frac{|a_{i,1}|-1}{\lambda_{c,i}}\geq\max_i\frac{|a_{i,1}|-1}{\max_j\lambda_{c,j}}$ and $\min_i\frac{|a_{i,1}|+1}{\lambda_{c,i}}\leq\min_i\frac{|a_{i,1}|+1}{\min_j\lambda_{c,j}}$, together yielding
		\begin{equation*}
		-\gamma_c-1 < \gamma_c\min_i|\det(A-\lambda_{p,i}A_p)|-\max_i|\det(A-\lambda_{p,i}A_p)|.  
		\end{equation*}
		Then, the two conditions above give \eqref{eq:consensusability_necessary}.
		$\hfill \triangleleft$
	\end{pf}
	
	\begin{remark}\label{rem:consensusability}
		We would like to emphasize that the condition in Proposition~\ref{prop:consensusability_necessary} is \textit{necessary}, meaning that \eqref{eq:consensusability_necessary} holds if the LIMAS in \eqref{eq:LIMAS_dyn} is consensusable. Thus, it can be utilized to verify that a LIMAS is not consensusable, in which case, there does not exist a static feedback gain $K$ that makes \eqref{eq:delta_dyn} stable. In order to show the role of physical coupling, we look at the extreme case of $A=0$, for which, \eqref{eq:consensusability_necessary} reads as $$\left|\det(A_p)\right|\left|\gamma_c\lambda_{p,min}^n-\lambda_{p,max}^n\right|<\gamma_c+1.$$ This condition is satisfied for small $|\det(A_p)|$, i.e., \textit{weak} physical coupling between subsystems. We also note that the condition given in Proposition~\ref{prop:consensusability_necessary} recovers the necessary condition in \cite{you2011} in the case of physically decoupled systems, i.e., when $A_p$=0.
	\end{remark}

	\subsection{First-order subsystems}\label{subsec:ScalarDynamics}
	
	As a simplified scenario, here we investigate the consensusability of LIMAS with scalar subsystem dynamics, i.e., $A,B,K \in \RR$, and $n=1$. In this case, the matrices $A$, $B$, and $K$ in~\eqref{eq:LIMAS_dyn} are denoted as $a$, $b$, and $k$, respectively. The matrix $A_p$ is omitted in the scalar case, since it can be lumped into $a_{ij}$s. Although this scenario is a subset of the higher-order dynamics in \ref{subsec:VectorDynamics}, we do not need the Assumptions~\ref{ass:lapl_commute} and \ref{ass:A_and_Ap} in the analysis of this section, which makes the results more general. The dynamics of the LIMAS with scalar subsystem dynamics is rewritten as 
	\begin{equation}\label{eq:LIMAS_dyn_scalar}
	x^+ = \left(a\I_N-\Lcal_p+bk\Lcal_c\right)x.
	\end{equation} 
	
	We note that, for scalar systems, Assumption~\ref{ass:controllability} corresponds to $b\neq 0$. Therefore, throughout this section, we assume $b=1$ without loss of generality. 
	
	\begin{proposition}\label{prop:consensusability_scalar}
		Suppose that Assumption~\ref{ass:controllability} holds. Define $\Delta_p \triangleq \lambda_{p,max}-\lambda_{p,min}\geq0$
		. Then, the scalar system \eqref{eq:LIMAS_dyn_scalar} is consensusable if one of the following conditions hold:
		\begin{enumerate}
			\item[C1.] $\lambda_{p,min}>a-1$ and $\left(\gamma_c-1\right)\left(1-a+\lambda_{p,min}\right)<\gamma_c\left(2-\Delta_p\right)$,
			\item[C2.] $\lambda_{p,max}<1+a$ and $\left(\gamma_c-1\right)\left(-1-a+\lambda_{p,max}\right)>\gamma_c\left(\Delta_p-2\right)$.
		\end{enumerate}
		Besides, the control gain $k$ can be selected as $k\in\{w|w\in \KK^+\cap\RR_{\geq0}\}$ if C1 is satisfied and as $k\in\{w|w\in \KK^-\cap\RR_{<0}\}$ if C2 is satisfied, where
		\begin{equation}\label{eq:K+K-}
		\begin{split}
		\KK^+ &= \left(\frac{-1-a+\lambda_{p,max}}{\lambda_{c,min}},\frac{1-a+\lambda_{p,min}}{\lambda_{c,max}}\right) \\
		\KK^- &=  \left(\frac{-1-a+\lambda_{p,max}}{\lambda_{c,max}},\frac{1-a+\lambda_{p,min}}{\lambda_{c,min}}\right)
		\end{split}.
		\end{equation}
		Moreover, a necessary condition for the consensusability of the scalar system \eqref{eq:LIMAS_dyn_scalar} is:
		\begin{equation}
		\left|\gamma_c\min_i|a-\lambda_{p,i}|-\max_i|a-\lambda_{p,i}|\right|<\gamma_c+1.
		\end{equation}
	\end{proposition}
	
	\begin{pf}
		We first note that Assumption~\ref{ass:controllability} is satisfied for $b=1$ in the scalar case. We then go on to realize that, in the scalar case, the dynamics of $\delta$ can be written as
		\begin{equation}\label{eq:delta_dyn_scalar}
			\delta^+ = \underbrace{\left(a\I_N-\Lcal_p+k\Lcal_c\right)}_{\tilde{A}}\delta.
		\end{equation}
		We note that $\tilde{A}$ is symmetric with an eigenpair $\{a,\1_N/\sqrt{N}\}$; therefore, for a unitary matrix $\Psi = [\1_N/\sqrt{N},\psi_2,\dots,\psi_N]$ the dynamics of $\tilde{\delta} \triangleq \Psi^{\top}\delta$ can be defined as
		\begin{equation}\label{eq:deltatilde_dyn_scalar}
			\tilde{\delta}^+ = \underbrace{\Psi^{\top}\tilde{A}\Psi}_{\tilde{\Lambda}}\tilde{\delta} = \left(a\I_N-\begin{bmatrix}
				0 & 0 \\
				0 & \tilde{\Lcal}_p
			\end{bmatrix}+k\begin{bmatrix}
				0 & 0 \\
				0 & \tilde{\Lcal}_c
			\end{bmatrix}\right)\tilde{\delta},
		\end{equation}
		where $\tilde{A}$ and $\tilde{\Lambda}$ are congruent to each other, and $\tilde{\Lcal}_p$ and $\tilde{\Lcal}_c$ are positive definite matrices with eigenvalues $\lambda_{p,i}$ and $\lambda_{c,i}$ for $i\in\{2,\dots,N\}$, respectively. By definition, $\tilde{\delta}_1 = \1_N^{\top}\delta/\sqrt{N} = 0$; thus, the stabilizability of \eqref{eq:delta_dyn_scalar} is equivalent to the existence of a scalar $k$ such that the matrix $a\I_{N-1}-\tilde{\Lcal}_p+k\tilde{\Lcal}_c$ is stable.
		
		Noticing that both $a\I_{N-1}-\tilde{\Lcal}_p$ and $k\tilde{\Lcal}_c$ are symmetric matrices, one can utilize the results on the eigenvalues of summation of Hermitian matrices~(\cite{fulton2000}) to show that
		\small
		\begin{equation*}\label{eq:scalar_eigenvaluebounds}
			\begin{split}
				\lambda_{max}\left(a\I_{N-1}-\tilde{\Lcal}_p+k\tilde{\Lcal}_c\right) &\leq \lambda_{max}\left(a\I_{N-1}-\tilde{\Lcal}_p\right) + \lambda_{max}\left(k\tilde{\Lcal}_c\right) \\
				\lambda_{min}\left(a\I_{N-1}-\tilde{\Lcal}_p+k\tilde{\Lcal}_c\right) &\geq \lambda_{min}\left(a\I_{N-1}-\tilde{\Lcal}_p\right) + \lambda_{min}\left(k\tilde{\Lcal}_c\right) \\
			\end{split}.
		\end{equation*}
		\normalsize
		
		Therefore, \eqref{eq:delta_dyn_scalar} is stabilizable if it is possible to find a $k \in \RR$ such that 
		\begin{equation*}
			\begin{split} 
				&a-\lambda_{p,min}+\lambda_{max}\left(k\tilde{\Lcal}_c\right)<1  \quad \text{and}\\
				& a-\lambda_{p,max}+\lambda_{min}\left(k\tilde{\Lcal}_c\right)>-1,
			\end{split}
		\end{equation*}
		which gives way to
		\begin{equation}\label{eq:ktilde_conditions}
			\begin{split}
				&-1-a+\lambda_{p,max}<\lambda_{min}\left(k\tilde{\Lcal}_c\right) \quad \text{and}\\
				&\lambda_{max}\left(k\tilde{\Lcal}_p\right)<1-a+\lambda_{p,min}
			\end{split}.
		\end{equation}
		
		Realizing that the sign of $k$ changes the expression of $\lambda_{min}(k\tilde{\Lcal}_c)$ and $\lambda_{max}(k\tilde{\Lcal}_c)$, we inspect the two cases $k\geq 0$ and $k<0$ separately. For the first case, we have $\lambda_{min}(k\tilde{\Lcal}_c) = k\lambda_{c,min}$ and $\lambda_{max}(k\tilde{\Lcal}_c) = k\lambda_{c,max}$ and the conditions above simplify to $k\in\KK^+\cap\RR_{\geq0}$, where $\KK^+$ is as given in \eqref{eq:K+K-}.
		This is possible if $\KK^+\neq \emptyset$ and $\KK^+\cap\RR_{\geq0}\neq\emptyset$, which translate into $$\left(\gamma_c-1\right)\left(1-a+\lambda_{p,min}\right)<\gamma_c\left(2-\Delta_p\right)$$ and $1-a+\lambda_{p,min}>0$, respectively. For the second case, where $k<0$, we have $\lambda_{min}(k\tilde{\Lcal}_c) = k\lambda_{c,max}$ and $\lambda_{max}(k\tilde{\Lcal}_c) = k\lambda_{c,min}$. Conditions in \eqref{eq:ktilde_conditions} are then written as $k\in\KK^-\cap\RR_{<0}$, where $\KK^-$ is as given in \eqref{eq:K+K-}.
		This is possible if $\KK^-\neq \emptyset$ and $\KK^-\cap\RR_{<0}\neq\emptyset$, which translate into $$\left(\gamma_c-1\right)\left(-1-a+\lambda_{p,max}\right)>\gamma_c\left(\Delta_p-2\right),$$ and $-1-a+\lambda_{p,max}<0$, respectively. The necessary condition given in the proposition is the restatement of the condition given in Proposition~\ref{prop:consensusability_necessary} for scalar subsystem dynamics.
		$\hfill \triangleleft$
	\end{pf}
	
	\begin{remark}\label{rem:consensusability_scalar}
		The results presented in Proposition~\ref{prop:consensusability_scalar} give valuable insight into the \textit{interplay} between the agent dynamics $a$, physical interconnection $\Gcal_p$, and cyber interconnection $\Gcal_c$. In particular, these conditions depend on the scalar $a$ as well as eigenvalues of physical and communication Laplacians and how much they are separated from each other. Firstly, condition C2 is satisfiable only if $a>-1$. Moreover, both conditions are possible only if $\Delta_p < 2$, i.e., when the difference between the largest and second smallest eigenvalues of $\Lcal_p$ is not greater than $2$. This is possible if $\lambda_{p,max}$ is small or close to $\lambda_{p,min}$. We also note that, given $\Delta_p<2$ and the first inequality of C1 is satisfied, $\gamma_c$ values satisfying the second inequality can always be found in a neighborhood of $\gamma_c=1$. Moreover, this also holds for condition C2, showing that having a communication graph whose Laplacian has eigenvalues that are close to each other is a favorable scenario. As in Remark~\ref{rmrk:consensusability_sufficient}, these observations also show that having \textit{weaker} physical interconnections or physical and communication graphs that allow for faster information diffusion  are favorable for consensusability, which agrees with the results in \cite{you2011}.
	\end{remark}
	
	\section{Simulation Results}\label{sec:SimulationResults}
	
	We now validate our technical results via computer simulations. For this, we are interested in a system with dynamics in \eqref{eq:Si_dyn}, equipped with a consensus-based controller as in \eqref{eq:Si_contr}. System matrices are chosen as:
	\begin{equation}\label{eq:simul_sys_matrices}
	\begin{split}
	A = \begin{bmatrix}
	1 & 2 \\ 
	0 & 1.5
	\end{bmatrix}, \quad B = \begin{bmatrix}
	0 \\
	1
	\end{bmatrix}, \quad A_p = 0.3A.
	\end{split}
	\end{equation}
	
	The topology of $\Gcal_p$ is as in Figure~\ref{fig:LIMAS_struct} with uniform edge weights of $0.1$. Consequently, to satisfy Assumption~\ref{ass:lapl_commute}, the communication graph is selected to be a complete graph with unit edge weights, i.e., $\Lcal_c=\I_N-\frac{1}{N}\1_N\1_N^{\top}$. 
	
\begin{figure}[t]
		\centering
		\includegraphics[width=0.45\textwidth]{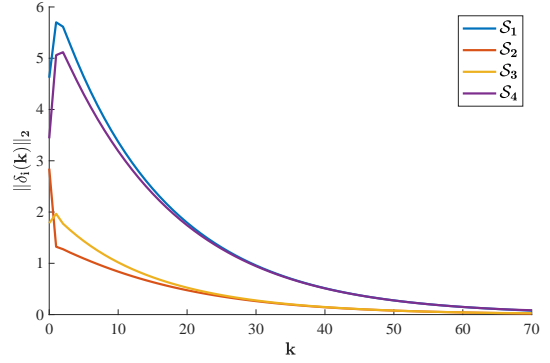}	\vspace{-.3cm}
		\caption{2-Norm of the deviation from consensus state for each subsystem.}
		\label{fig:delta_norm}
	\end{figure}

	Therefore, the LIMAS satisfies Assumptions~\ref{ass:lapl_commute}, \ref{ass:controllability}, and \ref{ass:A_and_Ap}, hence enabling the application of Proposition~\ref{prop:consensusability_sufficient} to see whether it is consensusable. Noting that $\lambda_{p} = \{0, 0.2, 0.2, 0.4\}$ and $\lambda_{c} = \{0,4,4,4\}$, one can see that the condition in \eqref{eq:consensusability_sufficient_critical} is satisfied. Thus, it is possible to solve \eqref{eq:MARE} through an LMI formulation and obtain the feedback gain $$K = -k(B^{\top}PB)^{-1}B^{\top}PA = \begin{bmatrix}
	0 & -0.3412
	\end{bmatrix}.$$ Since this LIMAS satisfies the sufficient condition in Proposition~\ref{prop:consensusability_sufficient}, it is consensusable and it will reach consensus with the static feedback gain $K$ calculated above. To verify this, we run a computer simulation on MATLAB for the closed-loop system with random initial states between $0$ and $10$. In Figure~\ref{fig:delta_norm}, we show that the 2-norm of the deviation from consensus state converges to zero for all subsystems $S_i$, i.e., consensus is asymptotically reached. 
	
	Having shown that this LIMAS is indeed consensusable and it reaches consensus from random initial states with a specific selection of matrix $K$, we verify that the necessary condition in Proposition~\ref{prop:consensusability_necessary} also holds. Observing that $\gamma_c=1$, it is straightforward to see that \eqref{eq:consensusability_necessary} holds, which confirms the result given in Proposition~\ref{prop:consensusability_necessary}.
	
	\section{Conclusion and Future Perspectives}\label{sec:ConclusionAndFuturePerspectives}
	
	In this paper, we have presented conditions on the consensusability of linear interconnected multi-agent systems, that are either necessary or sufficient. We analyzed the two cases where subsystems have first-order or higher-order dynamics. The latter class of LIMASs is more general, yet more challenging to analyze without additional assumptions. Instead, the analysis of the former class, albeit less general, leads to valuable insight into the consequences of having different interconnection and communication graphs. Our results show that, to achieve consensusability, it is favorable to have \textit{weaker} physical coupling, i.e., smaller $\alpha$ or smaller $\lambda_{p,max}$. In addition, smaller separation between the eigenvalues of the physical graph Laplacian, as well as the communication graph Laplacian, are also beneficial for consensusability. 
	
	The presented results in the case of vector dynamics rely heavily on the assumptions on the commutativity of Laplacians of physical and communication graphs, as well as on properties of the physical coupling. Therefore, in the future, we will work on eliminating these assumptions. Another possible research direction is to relate consensusability conditions to other properties of the two graphs, instead of only using the eigenvalue information. Analysis of the stability of a multi-layer consensus scheme, where there are multiple graphs through which subsystems are coupled with each other, is also a future perspective. 
	
	\bibliography{biblio_consensus}
	
\end{document}